\documentclass{aa}
\usepackage{graphicx}
\usepackage{natbib}
\usepackage{txfonts}
\bibpunct{(}{)}{;}{a}{}{,} % to follow the A&A style     
\begin{document}
\title{XMM-Newton spectral and timing analysis of the faint millisecond
  pulsars \object{PSR J0751+1807} and \object{PSR J1012+5307}  }

   \subtitle{}

   \author{N.A. Webb
          \inst{1}
          \and
          J.-F. Olive
          \inst{1}
          \and
          D. Barret
          \inst{1}
          \and
          M. Kramer
          \inst{2}
          \and
          I. Cognard
          \inst{3}
          \and
          O. L\"ohmer
          \inst{4}
          }

   \offprints{N.A. Webb, \email{Natalie.Webb@cesr.fr}}

   \institute{Centre d'Etude Spatiale des Rayonnements, 9 avenue du Colonel Roche, 31028 Toulouse Cedex 04, France 
   \and
   University of Manchester, Jodrell Bank Observatory, Macclesfield, Cheshire, SK11 9DL, UK 
   \and
   LPCE/CNRS Orleans, 3A avenue de la Recherche Scientifique, F-45071 Orleans Cedex 02, France
    \and
    Max-Planck-Institut f\"ur Radioastronomie, Auf dem H\"ugel 69, 53121 Bonn, Germany.
    }
   \date{Received  / Accepted }

   \abstract{We present XMM-Newton MOS imaging and PN timing data of
   the faint millisecond pulsars \object{PSR J0751+1807} and
   \object{PSR J1012+5307}.  We find 46 sources in the MOS field of
   view of \object{PSR J0751+1807} searching down to an unabsorbed
   flux limit of 3 $\times 10^{-15}$ ergs\ cm$^{-2}$ s$^{-1}$
   (0.2-10.0 keV).  We present, for the first time, the X-ray spectra
   of these two faint millisecond pulsars.  We find that a power law
   model best fits the spectrum of \object{PSR J0751+1807},
   $\Gamma$=1.59$\pm$0.20, with an unabsorbed flux of 4.4$\times
   10^{-14} {\rm ergs\ cm}^{-2} {\rm s}^{-1}$ (0.2-10.0 keV). A power
   law is also a good description of the spectrum of \object{PSR
   J1012+5307}, $\Gamma$=1.78$\pm$0.36, with an unabsorbed flux of
   1.2$\times 10^{-13} {\rm ergs\ cm}^{-2} {\rm s}^{-1}$ (0.2-10.0
   keV).  However, a blackbody model can not be excluded as the best
   fit to this data.  We present some evidence to suggest that both of
   these millisecond pulsars show pulsations in this X-ray band.  We
   find some evidence for a single broad X-ray pulse for \object{PSR
   J0751+1807} and we discuss the possibility that there are two
   pulses per spin period for \object{PSR J1012+5307}.

     }

\authorrunning{Webb et al.}
\titlerunning{XMM-Newton analysis of PSR J0751+1807 and PSR J1012+5307}

   \maketitle
%
%________________________________________________________________

\section{Introduction}

X-ray emission from millisecond pulsars (MSPs) is thought to be from:
charged relativistic particles accelerated in the pulsar magnetosphere
(non-thermal emission indicated by a hard power-law spectrum and sharp
pulsations); and/or thermal emission from hot polar caps; and/or
emission from a pulsar driven synchrotron nebula; or interaction of
relativistic pulsar winds with either a wind from a close companion
star or the companion star itself \cite[see e.g.][for a more thorough
review]{beck02}.  Until now, X-ray pulsations and spectra have often
not been observable for faint MSPs e.g. \object{PSR J0751+1807}
\citep{beck96} and \object{PSR J1012+5307} \citep{halp97}.  Thus it
has been difficult to discriminate between competing neutron star
models.  However, taking advantage of the large collecting area of
{\em XMM-Newton} \citep{jans01}, it is becoming possible to observe
not only the X-ray spectra but also put a limit on the presence of
X-ray pulsations of these faint MSPs.

We have observed two faint MSPs with {\em XMM-Newton}.  \object{PSR
J0751+1807} was first detected in an EGRET source error box, in
September 1993 \citep{lund93}, using the radio telescope at Arecibo.
\cite{lund95} combined the mass function, eccentricity, orbital size
and age of the pulsar, determined from radio data, to predict the
expected type of companion star to the millisecond pulsar.  They
proposed that the secondary star is a helium white dwarf, with a mass
between 0.12-0.6M$_\odot$, in a 6.3 hour orbit with the pulsar.  They
determined a 3.49 ms pulse period, but from the period derivatives,
the spin down energy indicated that the pulsar is not the source of
the $\gamma$-rays that were originally detected by EGRET.  \object{PSR
J0751+1807} was subsequently detected in the soft X-ray domain by
\cite{beck96}, using the ROSAT PSPC.  However, there were too
few counts detected to fit a spectrum or detect pulsations. Using the
HI survey of \cite{star92}, they deduced an interstellar absorption of
4$\times$ 10${\rm^{20}}$ cm${\rm^{-2}}$.  Then using their estimated
counts and assuming a powerlaw spectrum of $dN/dE \propto E^{-2.5}$,
they determined an unabsorbed flux of 1 $\times$ 10${\rm^{-13}}$
erg cm${\rm^{-2}}$ s${\rm^{-1}}$ (0.1-2.4 keV).  The corresponding
X-ray luminosity was then calculated to be L$_x$ = 4.7 $\times$
10${\rm^{31}}$ erg s${\rm^{-1}}$, for a distance of 2 kpc.  The
distance they used was calculated using the radio dispersion measure
and the model of \cite{tayl93} for the galactic distribution of free
electrons.

\object{PSR J1012+5307}, a 5.26 msec pulsar, was discovered by 
\cite{nica95} during a survey for short period pulsars conducted 
with the 76-m Lovell radio telescope.  From the dispersion measure
they found a distance of 0.52 kpc.  \cite{call98}, and references
therein, confirmed a white dwarf secondary of mass
0.16$\pm$0.02M$_\odot$ in a 14.5 hour circular orbit with the pulsar.
\cite{halp96} associated a faint (L$_x \approx $2.5$\times$ 10$^{30}$
ergs s${\rm^{-1}}$, 0.1-2.4 keV) X-ray source detected with the ROSAT
PSPC (80$\pm$24 photons), with the radio MSP \object{PSR J1012+5307}.
However, the number of photons was insufficient to determine a
spectrum or any pulsations.

In this work we present the X-ray spectrum of both \object{PSR
J0751+1807} and \object{PSR J1012+5307} for the first time.  We also
present some evidence for X-ray pulsation from both of these faint
pulsars and we compare their nature with other millisecond pulsars,
e.g. \object{PSR J0437-4715} \citep[][]{beck93,zavl98,zavl02}.

\section{Observations and data reduction}

\object{PSR J0751+1807} was observed by XMM-Newton on 2000 October 1.
The observations spanned 38 ksecs (MOS cameras) and 36.8 ksecs (PN
camera), but a solar flare affected approximately 8 ksecs of these
observations.  The second of our faint MSPs, \object{PSR J1012+5307},
was observed on 2001 April 19.  The MOS observations lasted 20.8 ksecs
and the PN observations 19.2 ksecs.  However the whole observation was
strongly affected by a solar flare.  The MOS data were reduced using
Version 5.4.1 of the {\it XMM-Newton} SAS (Science Analysis Software).
However, for the PN data we took advantage of the development track
version of the SAS.  Improvements have been made to the {\em oal} (ODF
(Observation Data File) access layer) task (version 3.106) to correct
for spurious and wrong values, premature increments, random jumps and
blocks of frames stemming from different quadrants in the timing data
in the PN auxillary file, as well as correcting properly for the
onboard delays \citep{kirs03}.  Indeed several spurious jumps were
corrected using this version.  We have verified that this version
improved our timing solution using the pulsar \object{PSR
J0218+4232} \citep[see][]{webb03}.

We employed the MOS cameras in the full frame mode, using a thin
filter \citep[see][]{turn01}.  The MOS data were reduced using
`emchain' with `embadpixfind' to detect the bad pixels.  The event
lists were filtered, so that 0-12 of the predefined patterns (single,
double, triple, and quadruple pixel events) were retained and the high
background periods were identified by defining a count rate threshold
above the low background rate and the periods of higher background
counts were then flagged in the event list.  We also filtered in
energy. We used the energy range 0.2-10.0 keV, as recommended in the
document `EPIC Status of Calibration and Data Analysis'
\citep{kirs02}.  The event lists from the two MOS cameras were merged,
to increase the signal-to-noise.

The PN camera was also used with a thin filter, but in timing mode
which has a timing resolution of 30$\mu$s \citep{stru01}.  The PN data
were reduced using the `epchain' of the SAS.  Again the event lists
were filtered, so that 0-4 of the predefined patterns (single and
double events) were retained, as these have the best energy
calibration.  We also filtered in energy.  The document `EPIC Status
of Calibration and Data Analysis' \citep{kirs02} recommends use of PN
timing data above 0.5 keV, to avoid increased noise.  We used the data
between 0.6-10.0 keV as this had the best signal-to-noise. The times
of the events were then converted from times expressed in the local
satellite frame to Barycentric Dynamical Time, using the task
`barycen' and the coordinates derived from observing the pulsar 4-8
times per month, over years, with the radio telescope at Nan\c{c}ay,
France, by one of us (IC), for \object{PSR J0751+1807}.  For
\object{PSR J1012+5307} we used the results from the high-precision
timing observations taken since 1996 October, using the 100-m
Effelsberg radio telescope and the 76-m Lovell telescope by two of us
(MK and OL).

\section{\object{PSR J0751+1807}}
\label{sec:0751}

\subsection{Spectral analysis}
\label{sec:0751spectra}

\begin{figure}
   \includegraphics[width=6cm,angle=270]{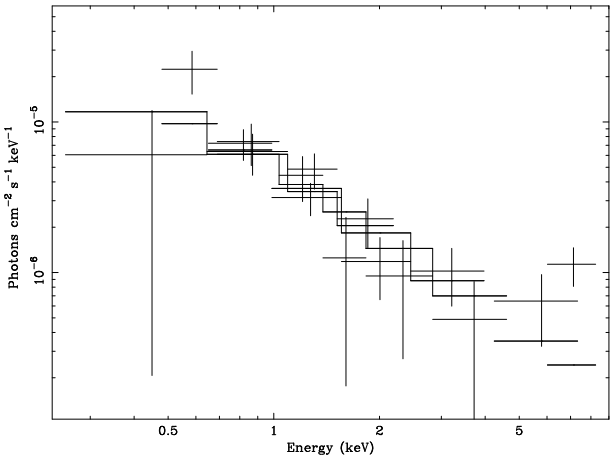} \caption{The
   combined MOS and PN spectrum of \object{PSR J0751+1807} fitted with
   a power law model.  The fit parameters can be found in
   Table~\ref{tab:src52specfits}.}
\label{fig:psr0751power}
\end{figure}

We extracted the MOS spectra of \object{PSR J0751+1807} using an
extraction radius of $\sim$1\arcmin\ and rebinned the data into 15 eV
bins.  We used a similar neighbouring surface, free from X-ray sources
to extract a background file.  We used the SAS tasks `rmfgen' and
`arfgen' to generate a `redistribution matrix file' and an `ancillary
response file'.  We binned up the data to contain at least 20
counts/bin.  The PN data were extracted in a similar way, following
the usual XMM-Newton timing data procedure.  The data in the RAWY
direction were binned into a single bin and the spectrum was extracted
using a rectangle of 1 $\times$ 3 pixels, which included all the
photons from the pulsar, see e.g. \cite{kust02}.  The background
spectrum was extracted from a similar neighbouring surface, free from
X-ray sources.  Again we used the SAS tasks `rmfgen' and `arfgen' to
generate a `redistribution matrix file' and an `ancillary response
file'.  We then used Xspec (Version 11.1.0) to fit the spectrum. We
tried to fit simple models to the combined PN and MOS spectra.  We
find the model fits as given in Table~\ref{tab:src52specfits} for the
spectrum between 0.2-10.0 keV, when the $N_H$ was frozen at 4 $\times
10^{20} {\rm cm}^{-2}$, \citep[see][]{beck96}.  We find that the best
fitting spectrum is a single power law, with a similar photon index to
the X-ray spectrum of other millisecond pulsars e.g. PSR B1821-24
\citep{sait97}.  The spectrum of \object{PSR J0751+1807}, plotted with
the power law fit, can be seen in Fig.~\ref{fig:psr0751power}.
Allowing the $N_H$ to vary, gives values compatible with the above
values.  We determine an unabsorbed flux of 4.4 $\times 10^{-14}\ {\rm
ergs\ cm}^{-2} {\rm s}^{-1}$ (0.2-10.0 keV).

\begin{table*}
\begin{minipage}{18cm}
\caption{The best model fits to the MOS spectra of the source 52, and the MOS and PN spectra of \object{PSR J0751+1807} and \object{PSR J1012+5307} .  For \object{PSR J0751+1807} the interstellar absorption ( N$_H$) was fixed at 4 $\times$ 10${\rm^{20}}$ cm$^{-2}$.  For \object{PSR J1012+5307} the N$_H$ was fixed at 7 $\times$ 10${\rm^{19}}$ cm$^{-2}$.  For source 52, z was fixed at 0.255.  All the errors shown are 1$\sigma$ confidence limits.  The fluxes and luminosities are given for the 0.2-10.0 keV band.}
\label{tab:src52specfits}
\begin{center}
\begin{tabular}{llcccccccc}
\hline
\hline
 Object & Spectral & N$_H$ (cm$^{-2}$) & kT & Photon & Abundance &
 $\chi^{\scriptscriptstyle 2}_{\scriptscriptstyle \nu}$ & dof & Flux
 ($\times 10^{-13}$) & Luminosity\\ & model & ($\times$
 10${\rm^{22}}$) & (keV) & Index & & & & (${\rm ergs\ cm}^{-2} {\rm
 s}^{-1}$) & (${\rm ergs\ s}^{-1}$)\\
\hline
{\bf \object{PSR J0751+1807}}  & Power law & 0.04 & - & 1.59$\pm$0.20 & - & 1.33 & 14 & 0.44 & 2.1$\times 10^{31}$ \\
 & Bremsstrahlung & 0.04 & 12.36$\pm$12.31 & - & & 1.52 & 14 & & \\
 & Blackbody & 0.04  & 0.32$\pm$0.04 & - & & 2.21 & 14 & &\\
\hline
{\bf Source 52}   & Power law & 0.13$\pm$0.06 & - & 1.46$\pm$0.13 & - & 0.86 & 32 & 2.0 & 1.9$\times 10^{43}$\\
 & Bremsstrahlung & 0.09$\pm$0.05 & 17.31$\pm$10.42 & - & & 0.87 & 32 & &\\
 & Raymond Smith & 0.07$\pm$0.04 & 17.63$\pm$12.78 & - & 0.16$\pm$1.20 & 0.89 & 31 & &\\
\hline
{\bf \object{PSR J1012+5307}} & Power law & 0.007 & - & 1.78$\pm$0.36 & - & 1.27 & 9 & 1.2 & 3.9$\times 10^{30}$\\
 & Bremsstrahlung & 0.007 & 1.95$\pm$1.45 & - & & 1.29 & 9 & &\\
 & Blackbody & 0.007  & 0.26$\pm$0.04 & - & & 1.37 & 9 & & \\
\hline
\end{tabular}
\end{center}
\end{minipage}
\end{table*}

\subsection{Timing analysis}
\label{sec:0751pntiming}

\begin{figure}
   \includegraphics[width=8cm]{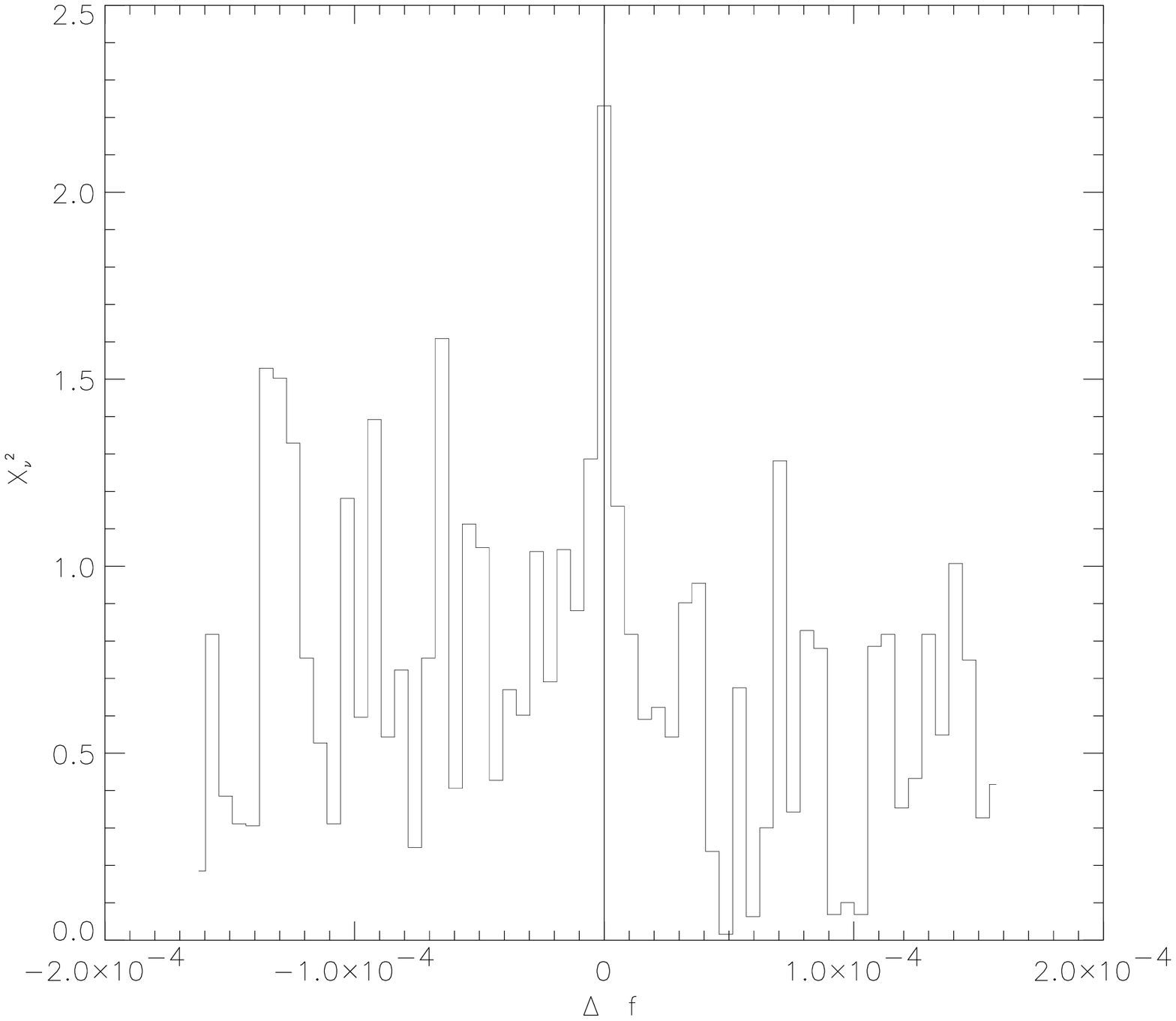}
   \caption{$\chi^{\scriptscriptstyle 2}_{\scriptscriptstyle \nu}$
   versus change in frequency from the expected pulsation frequency
   (shown as the solid vertical line at $\Delta$ f = 0.0) for
   \object{PSR J0751+1807}.} \label{fig:0751chisquare}
\end{figure}

\begin{figure}
   \includegraphics[width=8cm]{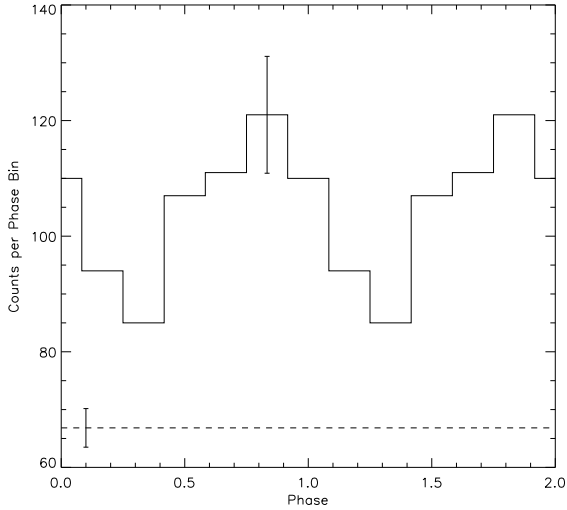} 
   \caption{Lightcurve
     of \object{PSR J0751+1807} folded on the radio ephemeris and binned
     into 6 bins, each of 0.58 msecs.  Two cycles are shown for clarity.
     A typical $\pm$1$\sigma$ error bar is shown.  The dashed line shows
     the background level, where the error bar represents the
     $\pm$1$\sigma$ error.}  
   \label{fig:0751foldedlc}
\end{figure}

We have reduced and analysed the timing data in the same way as
the timing analysis that we carried out on the MSP \object{PSR
J0218+4232} \citep{webb03}, which is a MSP that is known to show
pulsations in both the radio and X-rays.  We corrected the timing
data for the orbital movement of the pulsar and the data were folded
on the radio ephemeris, see Table~\ref{tab:0751parameters}, taking
into account the time-delays due to the orbital motion.  We used the
data between 0.6-7.0 keV, as we found that the majority of the
emission from \object{PSR J0751+1807} was in this energy band (see
Sect.~\ref{sec:0751spectra}) and thus the signal-to-noise in this band
was the best.  We tested the hypothesis that there is no pulsation in
the MSP \object{PSR J0751+1807}.  We searched frequencies at and
around the expected frequency and we found the largest peak in the
$\chi^{\scriptscriptstyle 2}_{\scriptscriptstyle \nu}$ versus change
in frequency from the expected frequency at the expected value, see
Fig~\ref{fig:0751chisquare} where we have taken the resolution of the
data (n) and plotted 4n bins in the range -1.75$\times 10^{-4}<
\Delta \rm f < 1.75\times 10^{-4}$.  Testing the significance of the
peak \citep{bucc85}, we find that it is significant at
 1.7$\sigma$. However, as this is the largest peak when searching 700
frequencies about the expected frequency and it falls at the expected
value, we tried to fold the data on this frequency.  The folded
lightcurve (0.6-7.0 keV), counts versus phase, is shown in
Fig~\ref{fig:0751foldedlc}. We find one broad pulse per period.
Fitting the lightcurve with a Lorentzian \citep[as][]{kuip02} we find
that the FWHM of the pulse is $\delta\phi_1$=0.311$\pm$0.1, centred at
phase $\phi_1$=0.38$\pm$0.04 (errors are 90\% confidence).  Fitting
with a Gaussian gives similar results. Using a Z$^{\scriptscriptstyle
2}_{\scriptscriptstyle 2}$ test \citep{bucc83}, which is independent
of binning, we determine a value of 5.5, which
corresponds to a probability that the pulse-phase distribution
deviates from a statistically flat distribution of 0.94.  The pulsed
percentage is 52$\pm$8\%.

\begin{table}[t]
\caption{Ephemeris of \object{PSR J0751+1807} from the Nan\c{c}ay radio timing data.}
\label{tab:0751parameters}
\begin{center}
\begin{tabular}{ll}
\hline
\hline
Parameter & Value \\
\hline
Right Ascension (J2000) & 07$^{\rm h}$ 51$^{\rm m}$ 09${\scriptstyle .}\hspace*{-0.05cm}^{\scriptstyle \rm s}$156312 \\
Declination (J2000) & 18$^{\circ}$ 07$^{'}$ 38${\scriptstyle .}\hspace*{-0.05cm}^{\scriptstyle ''}$590620 \\
Period (P)  & 0.003478770781560571 s \\
Period derivative (\.{P}) & 0.726912 $\times 10^{-20}$ s s$^{-1}$ \\
Second period derivative  (\"{P}) & 2.48928 $\times 10^{-30}$ s s$^{-2}$\\
Frequency ($\nu$) & 287.457858811 Hz \\
Frequency derivative (\.{\hspace*{-0.13cm}$\nu$}) & -6.0066207 $\times 10^{-16}$ Hz s$^{-1}$\\
Second frequency derivative (\"{\hspace*{-0.15cm}$\nu$})  & -2.0569423 $\times 10^{-25}$ Hz s$^{-2}$\\
Epoch of the period (MJD) & 49301.5 \\
Orbital period & 22735.664643 s \\
a.sin i & 0.39660728 \\
Eccentricity &   0.0000566981 \\
Time of ascending node (MJD) & 49460.430540 \\
\hline
\end{tabular}

\end{center}
\end{table}

\begin{figure}
    \hspace*{-0.8cm}\includegraphics[width=11cm]{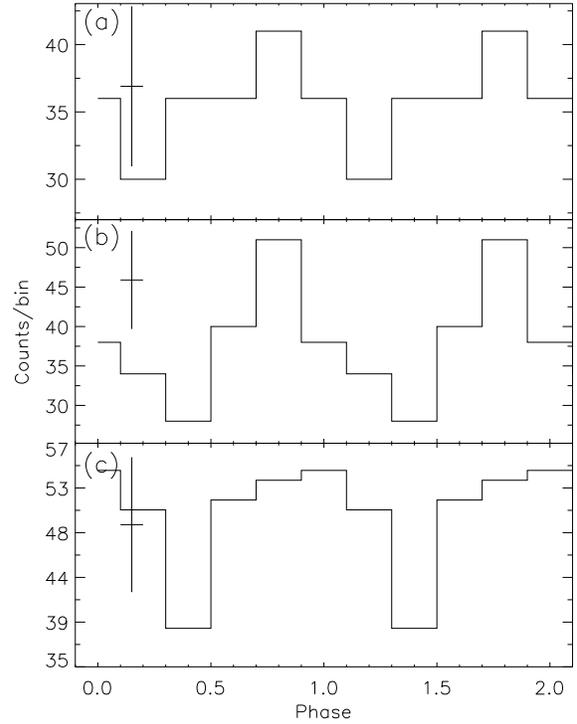}
   \caption{Lightcurves of \object{PSR J0751+1807} from different energy bands, folded on the
   radio ephemeris and binned into 5 bins.  Two cycles are shown for
   clarity.  (a) 0.6-1.0 keV; (b) 1.0-2.0 keV; (c) 2.0-7.0 keV. A
   typical $\pm$1$\sigma$ error bar is shown.  }
   \label{fig:0751energybands}
\end{figure}

Taking the 261$\pm$60 counts above the background in our observation,
we are also able to analyse the variations in the lightcurves from
different energy bands.  We chose the energy bands: 0.6-1.0 keV;
1.0-2.0 keV; and 2.0-7.0 keV.  The data were folded as before and
binned into 5 bins, where two cycles in phase, are presented for
clarity.  These lightcurves can be seen in
Fig.~\ref{fig:0751energybands}.  As can also be seen from the
spectral fitting, see Sect.~\ref{sec:0751spectra}, \object{PSR
J0751+1807} emits mostly at lower energies.  The lightcurves do not
vary dramatically, although the peak may be slightly narrower at lower
energies.  The pulsed percentage is possibly less at lower energies.  We
find 32$\pm16$\% for each of the energy bands: 0.6-1.0 keV; and
1.0-2.0 keV and 55$\pm23$\% for the 2.0-7.0 keV band.

\subsection{The MOS field of view}
\label{sec:0751mosfov}

\begin{figure}
   \includegraphics[width=8.8cm,angle=0]{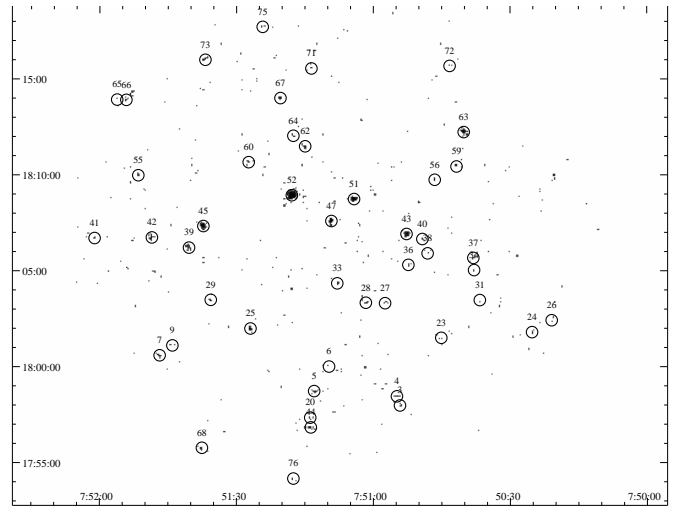} 
   \caption{The MOS
   field of view.  The sources detected with a likelihood
   $>$4.5$\sigma$ are encircled and labelled with a number.  These
   numbers correspond to the source identification numbers in
   Table~\ref{tab:0751fieldsources}.  \object{PSR J0751+1807} is
   labelled 47.}  \label{fig:0751field}
\end{figure}

We have detected 46 sources in total using the task `emldetect', with
a maximum detection likelihood threshold of 4.5$\sigma$ and ignoring
those sources not found on both cameras (unless they lay outside the
FOV or were in a chip gap). These sources can be seen in
Fig.~\ref{fig:0751field} and details such as their position, count
rates and likelihood of detection can be found in
Table~\ref{tab:0751fieldsources}. \object{PSR J0751+1807} is labelled
as source 47.  {\em ROSAT} detected 10 sources in the {\em XMM-Newton}
FOV \citep{beck96}.  We calculate an unabsorbed flux limit of our
observation of 3.0 $\times 10^{-15} {\rm ergs\ cm}^{-2} {\rm s}^{-1}$
(0.2-10.0 keV), using PIMMS (Mission Count Rate Simulator) Version
3.3a, with a power law, $\Gamma$ = 2 \citep[as][]{hasi01}.

\begin{table}[t]
\caption{X-ray sources in the \object{PSR J0751+1807} MOS field of view.  Information includes the source identification number, right ascension and declination of the source, count rate and detection likelihood.  The detection likelihood value is the value given by `emldetect' for detection in the two MOS cameras, corrected for the coding error in versions 4.11.15 and earlier.}
\label{tab:0751fieldsources}
\begin{center}
\begin{tabular}{lcccc}
\hline
\hline
Src & R.A. (2000) & dec. (2000) & Counts/s & Like- \\ ID & $^h$
\hspace*{3mm} $^m$ \hspace*{3mm} $^s$ & $^{\circ}$ \hspace*{3mm} '
\hspace*{3mm} '' & $\times10^{-3}$ & lihood \\
\hline
 3  & 7 50 54.1 & 17 57  59.3 &  5.6$\pm$0.8 &  62  \\
 4  & 7 50 54.8 & 17 58  27.9 &  5.6$\pm$0.8 &  63 \\
 5  & 7 51 12.9 & 17 58  44.3 &  4.4$\pm$0.6 &  53 \\
 13 & 7 51  9.7 & 18 0   1.04 &  3.3$\pm$0.5 &  40 \\
 17 & 7 51 46.8 & 18 0   36.0 &  6.4$\pm$0.8 &  99 \\
 19 & 7 51 44.0 & 18 1   7.05 &  3.8$\pm$0.7 &  32 \\
 20 & 7 51 13.8 & 17 57  21.3 &  3.7$\pm$0.7 &  23\\
 23 & 7 50 45.1 & 18 1   30.8 & 3.5$\pm$0.6 &   31 \\
 24 & 7 50 25.2 & 18 1   48.4 & 6.6$\pm$1.1 &   40 \\
 25 & 7 51 26.9 & 18 2   0.08 & 6.4$\pm$0.6 &   134 \\
 26 & 7 50 20.9 & 18 2   25.6 & 5.3$\pm$1.0 &   34 \\
 27 & 7 50 57.4 & 18 3   19.2 & 2.8$\pm$0.4 &   39 \\
 28 & 7 51  1.6 & 18 3   20.7 & 3.7$\pm$0.5 &   65 \\
 29 & 7 51 35.6 & 18 3   29.6 & 4.9$\pm$0.6 &   64 \\
 31 & 7 50 36.6 & 18 3   28.8 & 3.0$\pm$0.7 &   18 \\
 33 & 7 51  7.9 & 18 4   21.4 & 3.6$\pm$0.4 &   67 \\
 34 & 7 50 37.9 & 18 5    2.7 & 3.3$\pm$0.6 &   28 \\
 36 & 7 50 52.3 & 18 5   19.1 & 2.6$\pm$0.4 &   32 \\
 37 & 7 50 38.0 & 18 5   40.8 & 4.0$\pm$0.6 &   48 \\
 38 & 7 50 48.1 & 18 5   55.1 & 3.3$\pm$0.5 &   60 \\
 39 & 7 51 40.4 & 18 6   12.9 & 9.6$\pm$1.1 &   106 \\
 40 & 7 50 49.2 & 18 6   39.9 & 2.8$\pm$0.4 &   43 \\
 41 & 7 52  1.1 & 18 6   42.6 & 5.3$\pm$0.9 &   38 \\
 42 & 7 51 48.5 & 18 6   44.6 & 8.3$\pm$0.8 &   141 \\
 43 & 7 50 52.7 & 18 6   55.7 & 6.9$\pm$0.6 &   202 \\
 44 & 7 51 13.6 & 17 56  50.7 & 6.1$\pm$0.9 &   56 \\
 45 & 7 51 37.2 & 18 7   20.4 & 9.6$\pm$0.7 &   255 \\
 47 & 7 51  9.1 & 18 7   36.3 & 8.9$\pm$0.6 &   327 \\
 51 & 7 51  4.2 & 18 8   45.0 & 12.6$\pm$0.7 &   466 \\
 52 & 7 51 17.8 & 18 8   56.4 & 36.3$\pm$1.2 &   2017 \\
 55 & 7 51 51.5 & 18 9   59.2 & 4.2$\pm$0.6 &   44 \\
 56 & 7 50 46.5 & 18 9   45.2 & 4.6$\pm$0.7 &   56 \\
 59 & 7 50 41.7 & 18 10  26.7 &  3.0$\pm$0.5 &  37 \\
 60 & 7 51 27.3 & 18 10  40.2 &  3.0$\pm$0.5 &  41 \\
 62 & 7 51 14.9 & 18 11  30.0 &  3.1$\pm$0.4 &  56 \\
 63 & 7 50 40.1 & 18 12  14.3 &  17.2$\pm$1.2 &  355 \\
 64 & 7 51 17.5 & 18 12   2.9 &  2.7$\pm$0.4 &  47 \\
 65 & 7 51 56.1 & 18 13  55.7 &  5.4$\pm$1.0 &  37 \\
 66 & 7 51 54.1 & 18 13  54.9 &  8.6$\pm$1.2 &  73 \\
 67 & 7 51 20.3 & 18 14   0.7 &  6.6$\pm$0.6 &  117 \\
 68 & 7 51 37.5 & 17 55  46.9 &  6.9$\pm$1.1 &  53 \\
 71 & 7 51 13.5 & 18 15  33.4 &  3.7$\pm$0.6 &  40 \\
 72 & 7 50 43.2 & 18 15  41.3 &  4.4$\pm$0.8 &  28 \\
 73 & 7 51 36.8 & 18 16   0.5 &  6.3$\pm$0.8 &  96\\
 75 & 7 51 24.2 & 18 17  43.5 &  5.2$\pm$0.8 &  38 \\
 76 & 7 51 17.5 & 17 54  10.7 &  6.1$\pm$1.0 &  41 \\
\hline
\end{tabular}

\end{center}
\end{table}

\begin{figure}
   \includegraphics[width=6cm,angle=270]{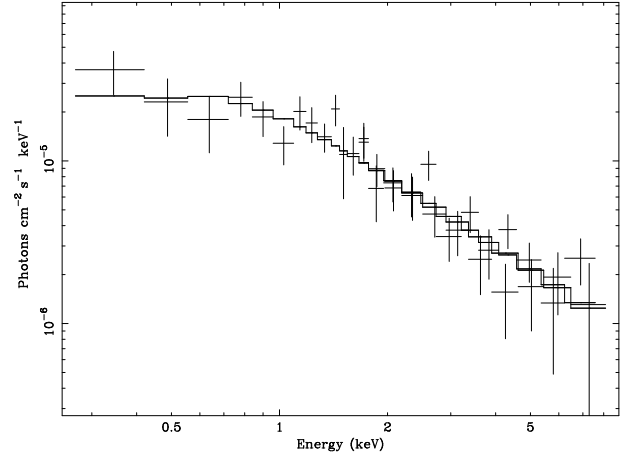} \caption{The
   spectrum of the source 52, in the \object{PSR
   J0751+1807} field of view, fitted with a power law model.  The fit
   parameters can be found in Table~\ref{tab:src52specfits}.}
\label{fig:src52power}
\end{figure}

The brightest source in the field of view is source 52, which was not
detected by {\em ROSAT}, where the deepest observation had a detection
limit of 1.5 $\times 10^{-14}\ {\rm ergs\ cm}^{-2} {\rm s}^{-1}$
(0.2-10.0 keV).  We detect this source with an unabsorbed flux of 2.0
$\times 10^{-13}\ {\rm ergs\ cm}^{-2} {\rm s}^{-1}$ (0.2-10.0 keV),
more than 13 times brighter than the {\em ROSAT} detection limit.
This source has been observed as a part of the AXIS (An XMM-Newton
International Survey) project \citep[see
http://www.ifca.unican.es/$\sim$xray/AXIS/ and also][]{barc02}.  The
source identified as the optical counterpart of our source 52 in the
i-band photometry is elongated and the spectrum shows calcium H and K
lines and the G-band.  From the spectral line shifts, a redshift of
z=0.255 has been determined by the AXIS group and they propose that
this source could be a galaxy.  We extracted the MOS spectra of source
52 in the same way as described in Section~\ref{sec:0751spectra}.  We
have fitted the spectrum with the extra-galactic origin in mind and
the results can be found in Table~\ref{tab:src52specfits}.  The data
fitted with a power law model can be found in
Fig.~\ref{fig:src52power}.  We investigated both models with the
redshift frozen at 0.255 and also models with the redshift as a free
parameter.  We found that the fits for the frozen and the variable
redshift (z) were the same within the error bars, hence we present
only one set of results, those where the redshift was fixed at 0.255.
We find a reasonably hard X-ray spectrum, which can not be fitted with
a black body.  We have also investigated a temporal variation, binning
the data into 1ks and 5ks bins.  However, we find no significant
variation during our observation.  From the X-ray spectral fitting and
short time scale temporal analysis, we can not support nor disregard
the galaxy hypothesis.

\section{\object{PSR J1012+5307}}

\subsection{Spectral analysis}
\label{sec:1012spectra}

Due to the observation of \object{PSR J1012+5307} being affected by a
solar flare during the whole of the observation, we detect only one
source in the whole of the field of view.  This source is at the
centre of the field of view, with the same coordinates as those of
\object{PSR J1012+5307} and is therefore supposed to be the MSP
\object{PSR J1012+5307}.  We have not carried out any filtering for
periods of higher background, as the whole observation would be
included.  Thus the signal to noise is poor for this observation.

\begin{figure}
   \includegraphics[width=6cm,angle=270]{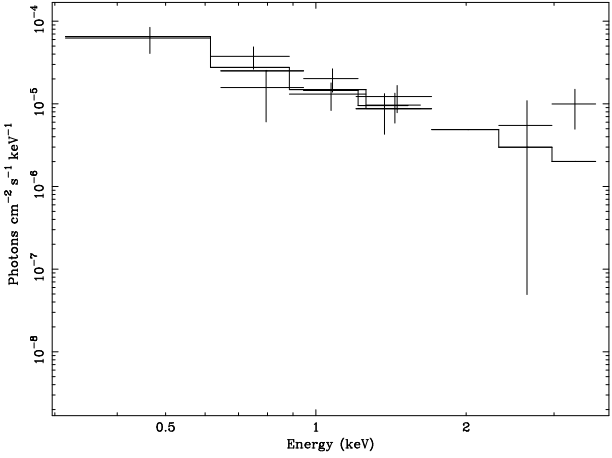}
   \caption{The combined MOS and PN spectrum of \object{PSR
   J1012+5307} fitted with a power law model.  The fit parameters can
   be found in Table~\ref{tab:src52specfits}.}
\label{fig:psr1012power}
\end{figure}

We extracted the spectra in the same way as described in
Sect.~\ref{sec:0751}.  We tried to fit simple models to the combined
PN and MOS spectra.  We find the model fits as given in
Table~\ref{tab:src52specfits} for the spectrum between 0.2-10.0 keV,
when the $N_H$ was frozen at 7 $\times 10^{19} {\rm cm}^{-2}$
\citep[see][]{snow94}.  We can not discriminate which of these 
fits is the best, however we present the spectrum of \object{PSR
J1012+5307} with a power law fit in Fig.~\ref{fig:psr1012power}.
Allowing the $N_H$ to vary, gives values compatible with the above
values.  We determine an unabsorbed flux of 1.2 $\times 10^{-13} {\rm
ergs\ cm}^{-2} {\rm s}^{-1}$ (0.2-10.0 keV).

\subsection{Timing analysis}
\label{sec:1012pntiming}

\begin{figure}
   \includegraphics[width=8cm]{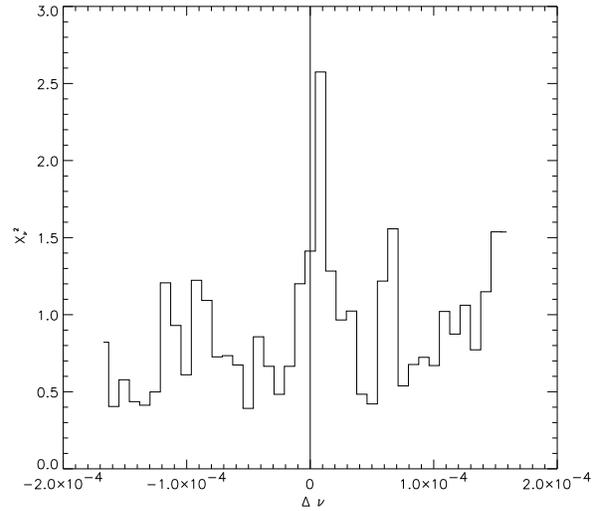}
   \caption{$\chi^{\scriptscriptstyle 2}_{\scriptscriptstyle \nu}$
   versus change in frequency from the expected pulsation frequency
   (shown as the solid vertical line at $\Delta$ f = 0.0) for
   \object{PSR J1012+5307}.} \label{fig:1012chisquare}
\end{figure}

\begin{figure}
   \hspace*{-1.5cm}\includegraphics[width=12cm]{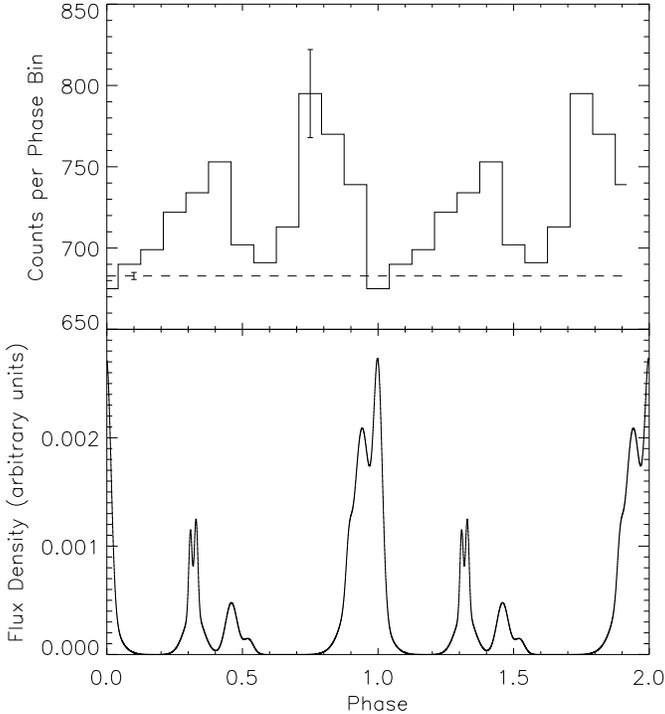}
   \caption{Upper panel: Lightcurve folded on the radio ephemeris and
   binned into 12 bins, each of 0.44 msecs.  Two cycles are shown for
   clarity.  A typical $\pm$1$\sigma$ error bar is shown.  The dashed
   line shows the background level, where the error bar represents the
   $\pm$1$\sigma$ error.  Lower panel: Radio profile
   from the Effelsberg and the Lovell radio timing data.  Again two
   cycles are shown for clarity.  } \label{fig:1012foldedlc}
\end{figure}

We corrected the timing data, as before, for the orbital
movement of the pulsar and the data were folded on new radio
ephemerides calculated to be correct for our observation (see
Table~\ref{tab:1012parameters}), taking into account the time-delays
due to the orbital motion.  We used the data between 0.6-5.0 keV, as
the signal-to-noise was best in this band.  Also from the spectral
fitting (see Sect.~\ref{sec:1012spectra}) we found that the majority
of the emission from \object{PSR J1012+5307} was in this energy band.
We tested the hypothesis that there is no pulsation in the MSP
\object{PSR J1012+5307}.  The largest peak in the
$\chi^{\scriptscriptstyle 2}_{\scriptscriptstyle \nu}$ versus change
in frequency from the expected frequency is at $\Delta \nu \simeq 8
\times 10^{-6} {\rm s}^{-1}$, see Fig~\ref{fig:1012chisquare}.  This implies a $\Delta {\rm f}$/f (=$\Delta {\rm
P}$/P) of $\sim$4 $\times 10^{-8}$, similar to the value that
we found for \object{PSR J0218+4232} \citep{webb03} ($\sim$1 $\times
10^{-8}$) and approaching the value of $< 10^{-9}$ determined from the
analysis of two revolutions of data of the MSP the \object{Crab}
\citep{kirs03,kirs03b}.  It
is also well inside the resolution of this dataset ($\sim$1/T$_{obs}$),
which is 7 $\times 10^{-5} {\rm s}^{-1}$, thus we can conclude that
the data reduction and analysis made to the dataset are reliable.  
Testing the significance of the peak
\citep{bucc85}, we find that it is significant at 
3$\sigma$. We folded the data on the radio period given in
Table~\ref{tab:1012parameters}.  The folded lightcurve (0.6-5.0 keV),
counts versus phase, is shown in the top panel of
Fig~\ref{fig:1012foldedlc}. We find some evidence for two pulses per
period. Fitting the lightcurve with two Lorentzians
\citep[as][]{kuip02} we find that the FWHM of the pulses is
$\delta\phi_1$=0.10$\pm$0.07, centred at phase
$\phi_1$=0.36$\pm$0.04 and $\delta\phi_2$=0.09$\pm$0.03, centred at
phase $\phi_2$=0.80$\pm$0.03.  Fitting with two Gaussians gives
similar results. Using a Z$^{\scriptscriptstyle 2}_{\scriptscriptstyle
4}$ test \citep{bucc83} on the X-ray data, we determine a value of
13, which corresponds to a probability that the pulse-phase
distribution deviates from a statistically flat distribution of 0.99.
We find a pulsed fraction of 77$\pm$13\%.  

In the lower panel in Fig~\ref{fig:1012foldedlc} we have plotted
the radio lightcurve from the timing observations taken using the
Effelsberg and Lovell telescopes.

%However, 
%Thus we have deemed it reasonable to try to
%fold the data on our determined frequency, which falls within the
%accuracy of PN timing data.  The folded lightcurve (0.6-5.0 keV),
%counts versus phase, is shown in the middle panel of
%Fig~\ref{fig:1012foldedlc}. We find two strong narrow pulses per
%period.  Fitting as before we find the FWHM of the pulses is
%$\delta\phi_1$=0.046$\pm$0.002, centred at phase
%$\phi_1$=0.27$\pm$0.05 and $\delta\phi_2$=0.034$\pm$0.004, centred at
%phase $\phi_2$=0.62$\pm$0.05.  Using a Z$^{\scriptscriptstyle
%2}_{\scriptscriptstyle 6}$ test, we determine a value of 21.2, which
%corresponds to a probability that the pulse-phase distribution
%deviates from a statistically flat distribution of 0.998 and with a
%Z$^{\scriptscriptstyle 2}_{\scriptscriptstyle 8}$ test, we determine a
%value of 45.2, which corresponds to a probability of 0.9999997.  We
%find a pulsed fraction of 77$\pm$9\%.

\begin{table}[t]
\caption{Ephemeris of \object{PSR J1012+5307} from the Effelsberg and the Lovell  radio timing data.  Errors on the last digits are shown in parenthesis after the values.}
\label{tab:1012parameters}
\begin{center}
\begin{tabular}{ll}
\hline
\hline
Parameter & Value \\
\hline
Right Ascension (J2000) & 10$^{\rm h}$ 12$^{\rm m}$ 33${\scriptstyle .}\hspace*{-0.05cm}^{\scriptscriptstyle \rm s}$43463677 \\
Declination (J2000) & 53$^{\circ}$ 07$^{'}$ 02${\scriptstyle .}\hspace*{-0.05cm}^{\scriptscriptstyle ''}$4965199 \\
Period (P)  & 0.00525605523 s \\
Period derivative (\.{P}) & 0.726912 $\times 10^{-20}$ s s$^{-1}$\\
Second period derivative  (\"{P}) & 2.48928 $\times 10^{-30}$ s s$^{-2}$\\
Frequency ($\nu$) & 190.267837551251880(508) Hz \\
Frequency derivative (\.{\hspace*{-0.13cm}$\nu$}) & -6.200023(202)$\times 10^{-16}$ Hz s$^{-1}$\\
Second frequency derivative (\"{\hspace*{-0.15cm}$\nu$})  & 1.871(294) $\times 10^{-27}$Hz s$^{-2}$\\
Epoch of the period (MJD) & 52018.324635450415 \\
Orbital period & 52243.7224464(17) s \\
a.sin i & 0.581817416(121) \\
Eccentricity &  $<$1.3 $\times 10^{-6}$ \\
Time of ascending node (MJD) & 52018.268144542(23) \\
\hline
\end{tabular}

\end{center}
\end{table}

\section{Discussion}

We have investigated the two faint millisecond pulsars \object{PSR
J0751+1807} and \object{PSR J1012+5307} in the X-ray band 0.2-10.0
keV, to ascertain the nature of the X-ray spectra. There are only a
handful of MSPs that have been seen to pulse in both the radio and in
X-rays to date \citep{beck02}, thus we have also investigated whether
these two MSPs show pulsations in X-rays.

\cite{sait97} showed that the MSP \object{PSR B1821-24} has a similar 
magnetic field value, at the light cylinder radius (B$_{\rm L}$), as
the \object{Crab pulsar}.  If this value can be taken to indicate the
high-energy magnetospheric activity, they state that one can expect to
see similar pulses from pulsars with a similar value of B$_{\rm L}$
and thus such pulsars are good candidates amongst the MSPs in which to
look for magnetospheric X-ray pulsation in high energy bands.  We have
calculated the B$_{\rm L}$ for \object{PSR J0751+1807} and \object{PSR
J1012+5307}, using the values for the magnetic field strength from
\cite{zhan00}.  We find 8.3$\times 10^4$ and 3.5$\times 10^4$ G
respectively. The value calculated for \object{PSR J0751+1807}
indicates that it is amongst the top 10 values for a millisecond
pulsar.  That of \object{PSR J1012+5307} places it in the top half.
Thus if a high B$_{\rm L}$ indicates high-energy magnetospheric
activity this adds support to the detection of X-ray pulsations in our
observation of \object{PSR J0751+1807} and possibly to that of
\object{PSR J1012+5307}.

We have also calculated the spin down energy (\.E) and the luminosity
in the 0.1-2.4 keV band of these two pulsars, to compare the results
with the correlation found between these two parameters by
\cite{beck97}.  \cite{beck97} suggest that pulsars that obey this 
relationship emit X-rays produced by magnetospheric emission,
originating from the co-rotating magnetosphere.  For \object{PSR
J0751+1807} we find log(\.E) of 34.35.  To calculate the luminosity,
we have used the distance calculated using the radio dispersion
measure and the model of \cite{tayl93}.  This gives a value of
log(L$_x$) of 31.0$\pm$0.2, which places the point close to the
expected value of 31.2 using the correlation proposed by
\cite{beck97}.  However, using the more recent model of
\cite{cord03a,cord03b} diminishes the distance by almost a factor 2,
to 1.1 kpc.  This gives a log(L$_x$) of 30.5$\pm$0.2 and thus
displaces the point further from the expected value.  For the MSP
\object{PSR J1012+5307} we find a log(\.E) of 34.2 and a log(L$_x$) of
30.4$\pm$0.3, where the expected log(L$_x$) is approximately 31.1.
This places the point quite some distance from the expected value,
which may indicate that the relationship is not a hard and fast rule.
Indeed \cite{poss02} found recently that when analysing pulsars in the
2.0-10.0 keV band, this relationship does not hold true.

\subsection{PSR J0751+1807}

We find that the best fitting model to the X-ray spectrum of
\object{PSR J0751+1807} is a power law.  This is indicative of a
magnetospheric origin of the X-ray emission.  We have shown evidence
that there appears to be a single broad pulse emitted from this MSP,
where the pulsation appears to change only slightly with increasing
energy (becomes slightly broader).  The observations also indicate
that the pulsed fraction possibly increases at higher energies. A
broad pulsation can be observed from pulsars showing hard
magnetospheric emission or soft thermal emission and thus we can not
yet discern the origin of the X-ray emission from \object{PSR
J0751+1807} with this data set.  Indeed this pulsar may be best fitted
by a multi-component model, as is the case with several other brighter
MSPs e.g. \object{PSR J0218+4232}, \citep{webb03} or \object{PSR
J0437-4715}, \citep{zavl98,zavl02}.  Longer observations of
\object{PSR J0751+1807} will help to distinguish the true nature of
the spectrum.

\subsection{PSR J1012+5307}

We can not discriminate which of the model fits to the MSP \object{PSR
J1012+5307} is the best, however we find that the single power law has
a similar photon index to the X-ray spectrum of \object{PSR
J0751+1807} (see Sect.~\ref{sec:0751spectra}), which could indicate
a magnetospheric origin of the X-ray emission.  However, the
temperature of the blackbody (1.9$\pm$0.5 $\times 10^6$ K) is
consistent with that emitted from the heated polar caps of a
millisecond pulsar \citep[10$^6$-10$^7$ K e.g.][and references
therein]{zhan03,zavl98}. Calculating the radius of the emission area
from the blackbody model fit, we find a radius of
0.05$\pm^{\scriptscriptstyle 0.01}_{\scriptscriptstyle 0.02}$ km, which
is  smaller than the expected radius of emission from polar
caps \citep[$\sim$1 km e.g.][and references therein]{zhan03,zavl98}.
\cite{zavl96} state however, that spectral fits with simplified
blackbody models can produce higher temperatures and smaller sizes due
to the fact that the X-ray spectra emerging from light-element
atmospheres are harder than blackbody spectra. Alternatively,
\cite{zavl98} and \cite{zavl02} suggest that the thermal emission 
can be from non uniform polar caps and we may therefore be seeing the
emission from the hotter central region of the caps.

Folding the timing data on the radio frequency, we find some evidence
for two pulses emitted from this MSP, separated by approximately 0.4
in phase.  This is similar to other millisecond pulsars,
e.g. \object{PSR J0218+4232} \citep{webb03} and \object{PSR B1821-24}
\citep{sait97}.

\section{Conclusions}

We have presented XMM-Newton data of the faint millisecond pulsars
\object{PSR J0751+1807} and \object{PSR J1012+5307}.  Both of these
pulsars have a reasonably large magnetic field at the light cylinder
radius, which could indicate that both of these MSPs should show
pulsations in X-rays.  We present for the first time the X-ray spectra
of these two faint millisecond pulsars.  We find that a power law
model best fits the spectrum of \object{PSR J0751+1807},
$\Gamma$=1.59$\pm$0.20, with an unabsorbed flux of 4.4 $\times
10^{-14} {\rm ergs\ cm}^{-2} {\rm s}^{-1}$ (0.2-10.0 keV). A power law
is also a good description of the spectrum of \object{PSR J1012+5307},
$\Gamma$=1.78$\pm$0.36, with an unabsorbed flux of 1.2 $\times
10^{-13} {\rm ergs\ cm}^{-2} {\rm s}^{-1}$ (0.2-10.0 keV).  However, a
blackbody model can not be excluded as the best fit to the data.  We
have also shown evidence to suggest that both of these MSPs may
show X-ray pulsations.  \object{PSR J0751+1807} appears to show a
single pulse, whereas \object{PSR J1012+5307} may show some evidence
for two pulses per pulse period.

\begin{acknowledgements}

We wish to thank A. Marcowith for his advice pertaining to the work in
this manuscript.  This article was based on observations obtained with
XMM-Newton, an ESA science mission with instruments and contributions
directly funded by ESA Member States and NASA.  The authors also
acknowledge the CNES for its support in this research.

\end{acknowledgements}

\end{document}